\documentclass[aps,prl,preprint]{revtex4} 
\usepackage{graphicx}

\begin{document}

\title{Wave interference effect on polymer microstadium laser}

\author{W. Fang, H. Cao}
 
\address{Department of Physics and Astronomy, Northwestern University, Evanston, IL 60208.}

\begin{abstract}
We investigate the lasing modes in fully chaotic polymer microstadiums under optical pumping. The lasing modes are regularly spaced in frequency, and their amplitudes oscillate with frequency. Our numerical simulations reveal that the lasing modes are multi-orbit scar modes. The interference of partial waves propagating along the constituent orbits results in local maxima of quality factor at certain frequencies. The observed modulation of lasing mode amplitude with frequency results from the variation of quality factor, which provides the direct evidence of wave interference effect in open chaotic microcavities. 
  
\end{abstract}

\maketitle

Over the past two decades, there have been many developments in microcavity lasers. Compared to inorganic semiconductor microlasers, polymer microlasers have the advantages of low fabrication cost and great substrate compatibility. Various shapes of microcavities have been fabricated with polymers and organic materials, e.g., circular disk, ring, ellipse, quadrupole, hexadecapole, spiral, stadium \cite{kuwata,frolov,polson,schwefel,ben,leb}. Among them, the two-dimensional (2D) stadium-shaped cavity is fully chaotic in the sense that no stable periodic orbit exists in the cavity. However, a dense set of unstable periodic orbits (UPOs) are embedded in the sea of chaotic orbits. Although the UPOs have zero measure in the classical ray dynamics, in wave mechanics they manifest themselves in terms of scar modes. Lasing oscillation has been realized in both scar modes and chaotic modes of GaAs microstadiums \cite{haraPRE,haraPRL,fangAPL07}. Since the refractive index of polymer is usually lower than that of GaAs, light confinement in a polymer stadium is expected to be weaker than that in a GaAs stadium. However, recent study shows that the stadium-shaped polymer microlaser has high efficiency and well-defined lasing threshold \cite{leb}. The lasing modes are regularly spaced in frequency, and the laser output is highly directional. Such good behavior of a chaotic microlaser is surprising. In order to understand it, we must know what the lasing modes are.  

In this letter, we investigate both experimentally and theoretically the lasing modes in polymer microstadiums with various shapes. Under optical pumping, we observe multiple groups of lasing modes with constant frequency spacing. The amplitudes of lasing modes oscillate with frequency.  Our numerical simulations reveal that the lasing modes are scar modes consisting of multiple UPOs. The interference of partial waves propagating along the constituent orbits maximize the optical confinement in the stadium at certain frequencies. The observed modulation of lasing mode amplitude with frequency results from the variation of quality factor.  

We fabricated stadium-shaped microcavities with a photoresist made of cresol novolak resin. Rhodamine 640 perchlorate dye was dissolved into the photoresist and served as the gain medium. The photoresist was spin-coated on a 500nm thick silica layer on top of a silicon substrate. The film thickness was $1.2 \mu$m. A set of microstadiums with the same area but different deformation were patterned by photolithography. The deformation $\epsilon$ is defined as the ratio between the length of the straight segment connecting the two half circles and the diameter of the half circle. 

In the lasing experiment, the Rhodamine 640 molecules were optically excited by the second harmonics of a Nd:YAG laser at $\lambda$ = 532nm. The pump pulse was 30ps long and the  repetition rate was 10Hz. An objective lens focused the pump beam onto a single stadium at normal incidence. The pump spot was larger than the stadium to ensure uniform pumping across the stadium. The emission was collected from the side of the stadium by another lens, and spectrally analyzed by a spectrometer with the resolution $\sim$ 0.1nm.   

We observed multi-mode lasing in microstadiums of $\epsilon$ ranging from 0.1 to 1.6. Figure 1 shows the lasing spectra of ten stadiums with $\epsilon$ between 0.2 and 1.1. The area of the stadiums is kept at $\sim 2000 \mu$m$^2$. The incident pump pulse energy is approximately 0.8$\mu$J. The spectra exhibit regularly-spaced lasing peaks. As given in Fig. 1, the frequency spacing of adjacent lasing peaks varies with $\epsilon$. At some deformation, e.g. $\epsilon = 1.1$, the lasing spectrum consists of several groups of lasing peaks. Although it is constant within each group, the peak spacing in different group may not be the same. Figure 1 also shows that the variation of lasing peak height with frequency changes from stadium to stadium. If all the lasing modes in a stadium had similar quality ($Q$) factors, their amplitudes would be determined by the optical gain at their frequencies. Since the gain spectrum is identical for all the stadiums, the different amplitude modulation of lasing modes indicate their quality factors differ significantly. In fact, the variation of lasing mode intensities reflects the change of their quality factors.    

To understand the experimental results, we carried out numerical simulations using the finite-difference time-domain (FDTD) method. To identify the lasing modes, we calculate the high-quality modes in 2D stadiums of size parameters identical to the measured ones. Details of our numerical method can be found in Ref.\cite{fang}. For the stadium with $\epsilon = 1.1$, we find four groups of high-$Q$ modes in the frequency range close to the experimental value. Figure 2 plots the $Q$ values of these modes versus their frequencies. Within each group the modes are equally spaced in frequency. The mode spacing $\Delta \nu$ of group I is the same as that of group II, while $\Delta \nu$ of group III is identical to that of group IV. From each group we select one mode whose $Q$ value is at or close to the maximum and plot its spatial intensity distribution in Fig. 3. The modes in groups I and II have the same spatial profiles, indicating they are the same type of modes. The modes in groups III and IV belong to a different type of modes as their spatial profiles are distinct from those in groups I and II.

The spatial profiles of the high-$Q$ modes reveal that they are not chaotic modes but scar modes \cite{fang,lee}. To find out the classical ray trajectories that these modes correspond to, we calculated the Husimi phase-space projection of these mode from the electric field distribution along the stadium boundary. Figure 4 shows the results for one mode in group I/II and another one in group III/IV. The mode in group I/II consists mainly of three  types of UPOs which are plotted in the inset of Fig. 4(a). Figure 4(b) reveals that the mode in group III/V contains mostly of two UPOs. The $Q$ variation of modes in group I/II or III/IV is attributed to interference of waves propagating along the constituent UPOs \cite{fangAPL07,fang,fukuJLT,fukuPRA}. The interference effect is sensitive to the relative phase of waves traveling in different orbits, which depends on frequency. At certain frequency, constructive interference may minimize light leakage out of the cavity, thus maximizing the quality factor. This leads to local maxima of $Q$ values for modes in groups I/II and III/IV, as shown in Fig. 2.

When a stadium is uniformly pumped, the lasing modes correspond to the high-$Q$ modes whose frequencies fall inside the gain spectrum. Typically the higher quality factor results in larger amplitude of a lasing mode. For the stadium of $\epsilon$ = 1.1, our simulation shows there is no other groups of high-$Q$ modes near the Rhodomaine 640 gain spectrum except the four shown in Fig. 2. We believe the three groups of lasing modes observed experimentally for$\epsilon$ = 1.1 in Fig. 1 correspond to the three groups of high-$Q$ modes at lower frequency in Fig. 2. The fourth group at higher frequency could not lase, because it falls outside the gain spectrum. The slight deviations of the calculated mode frequencies and the frequency spacings from the experimental values are attributed to the error in determining the refractive index of the dye-doped polymer near the resonance as well as the refractive index change induced by intense pumping in the lasing experiment. Therefore, the two groups of lasing modes with identical frequency spacing $\delta$ = 1.03THz belong to the same type of scar mode.  The spectral modulation of lasing mode amplitude in Fig. 1 is due to the variation of quality factor, which results from the interference of partial waves in the multi-orbit scar modes. 

In summary, we studied the lasing modes in polymer microstadiums with various deformations. At some deformations, we observed multiple groups of lasing modes with constant frequency spacing. Instead of following the gain spectrum, the amplitudes of lasing modes oscillate with frequency.  Our numerical simulations reveal that the lasing modes are high-quality scar modes consisting of multiple UPOs. The interference of partial waves propagating along the constituent orbits results in local maxima of quality factor at certain frequencies. The observed modulation of lasing mode amplitude with frequency results from the variation of quality factor. Our experimental results provide the direct evidence of wave interference effect on resonant modes in fully chaotic microcavities. 
   
We acknowledge Prof. J. Zyss and M. Lebental for stimulating discussions. This work is supported by NIST under the Grant No. 70NANB6H6162 and by NSF/MRSEC under the Grant No. DMR-00706097.

\pagebreak

\begin{figure}
\includegraphics[width=12cm]{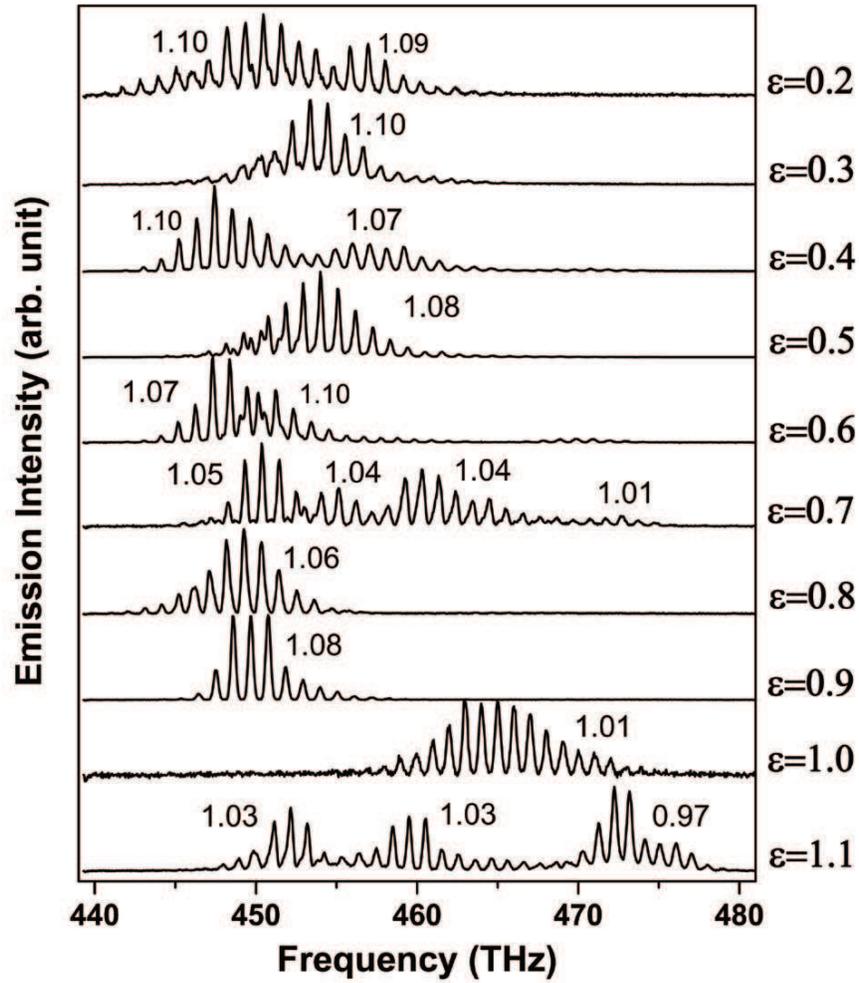}
\caption{\label{fig1}Measured spectra of laser emission from ten polymer stadiums. The deformation $\epsilon$ varies from 0.2 to 1.1. The stadium area is fixed at $\sim 2000 \mu$m$^2$. The incident pump pulse energy is approximately 0.8$\mu$J. The constant frequency spacing of lasing modes within each group is given in the unit of THz.}
\end{figure}

\begin{figure}
\includegraphics[width=12cm]{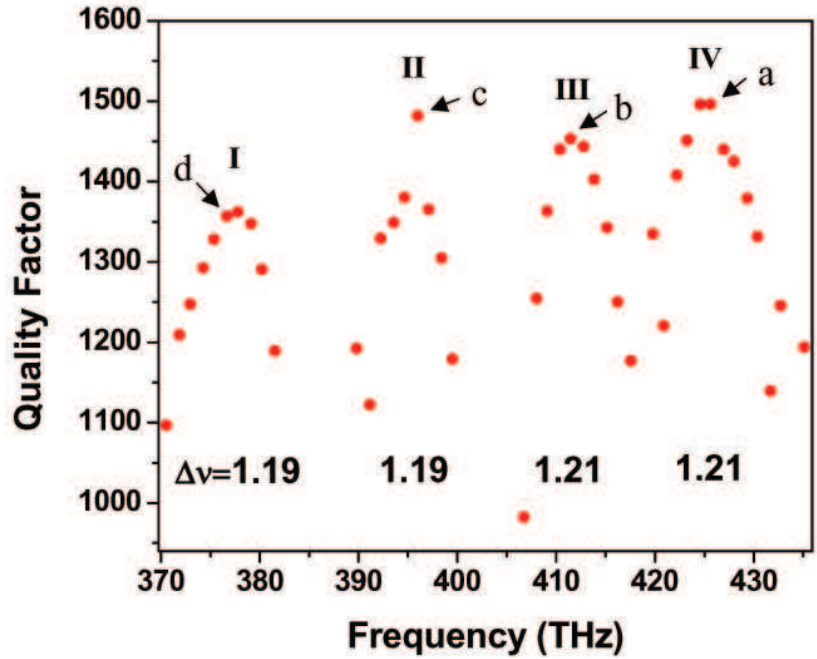}
\caption{\label{fig2}Calculated $Q$ values of the high-quality modes in a stadium with $\epsilon=1.1$ and area $\simeq 2000 \mu$m$^2$ versus the mode frequencies. The frequency spacing $\Delta \nu$ of adjacent modes in each of the four groups is given in the unit of THz.}
\end{figure}

\begin{figure}
\includegraphics[width=12cm]{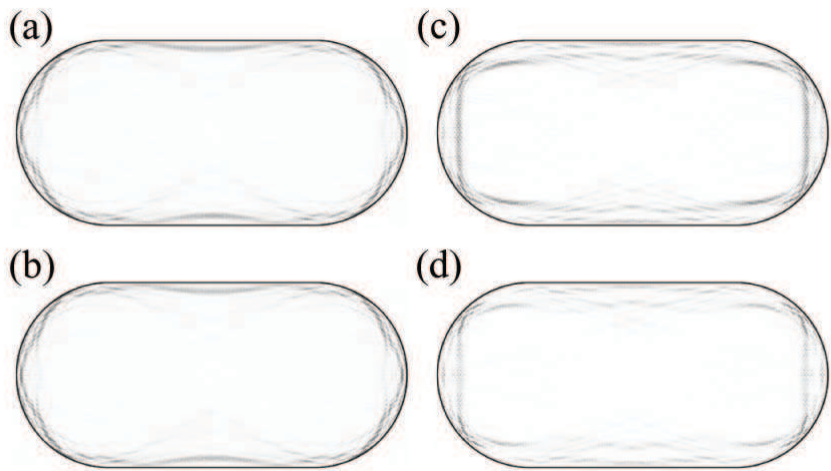}
\caption{\label{fig3}Calculated spatial intensity distribution for the four modes marked in Fig. 2.}
\end{figure}

\begin{figure}
\includegraphics[width=12cm]{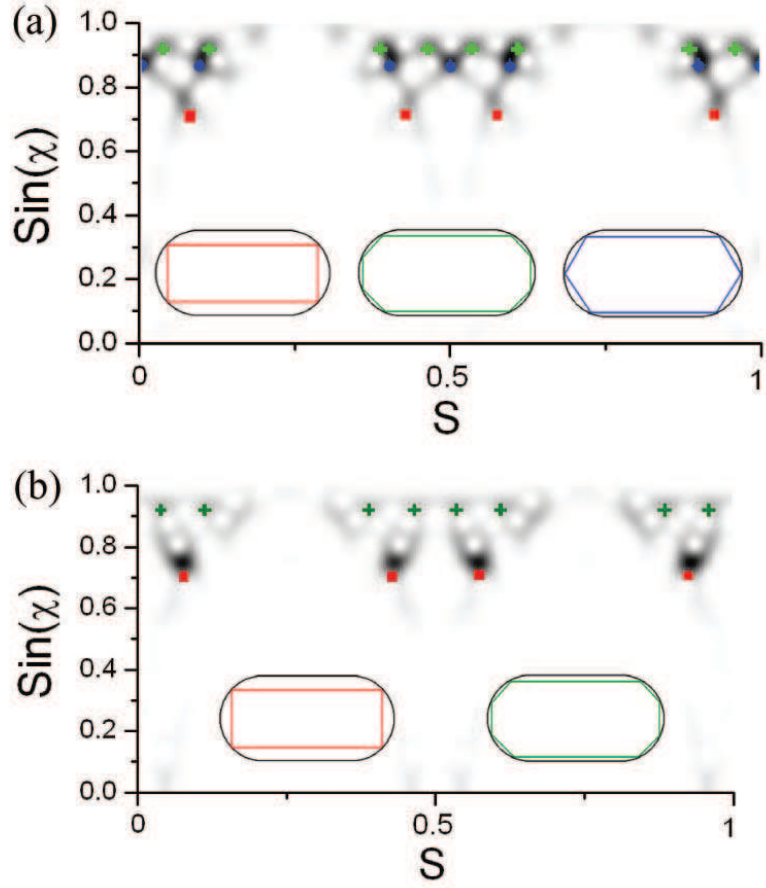}
\caption{\label{fig4}(a) and (b) are the Husimi phase space projection of the modes marked b and d in Fig. 2. The horizontal coordinate $S$ represents the length along the stadium boundary from the rightmost point, normalized by the stadium perimeter. The vertical axis corresponds to ${\rm sin} \chi$, where $\chi$ is the incident angle on the stadium boundary. Solid squares, circles and crosses mark the positions of the constituent UPOs plotted in the inset.}
\end{figure}

\end{document}